\documentclass[aps,prb,twocolumn,showpacs,superscriptaddress,floatfix,nofootinbib]{revtex4}
\usepackage{amsfonts,hyperref}

\usepackage{graphicx}
\usepackage{amsmath}
\usepackage{helvet} 
\usepackage{amssymb}

\newcommand{\be}{\begin{equation}}
\newcommand{\ee}{\end{equation}}
\newcommand{\bea}{\begin{eqnarray}}
\newcommand{\eea}{\end{eqnarray}}

\newcommand{\hc}{\hat c}

\begin{document}

\title{
Simulation of fermionic lattice models in two dimensions with Projected Entangled-Pair States: Next-nearest neighbor Hamiltonians
}
\author{Philippe Corboz}
\affiliation{School of Mathematics and Physics, The University of
Queensland, QLD 4072, Australia} 
\author{Jacob Jordan}
\affiliation{School of Mathematics and Physics, The University of
Queensland, QLD 4072, Australia} 
\author{Guifr\'e Vidal} 
\affiliation{School of Mathematics and Physics, The University of
Queensland, QLD 4072, Australia} 

\date{\today}

\begin{abstract}
In a recent contribution [Phys. Rev. B 81, 165104 (2010)] fermionic Projected Entangled-Pair States (PEPS) were used to approximate the ground state of free and interacting spinless fermion models, as well as the $t$-$J$ model. This paper revisits these three models in the presence of an additional next-nearest hopping amplitude in the Hamiltonian. First we explain how to account for next-nearest neighbor Hamiltonian terms in the context of fermionic PEPS algorithms based on simulating time evolution. Then we present benchmark calculations for the three models of fermions, and compare our results against analytical, mean-field, and variational Monte Carlo results, respectively. Consistent with previous computations restricted to nearest-neighbor Hamiltonians, we systematically obtain more accurate (or better converged) results for gapped phases than for gapless ones. 
\end{abstract}


\maketitle

\section{Introduction} 
Several recent papers have proposed and explored the use of tensor networks to simulate fermionic lattice models in two spatial dimensions,\cite{Corboz09, Kraus09, Pineda09, Corboz09b, Barthel09, Shi09, Corboz10b, Pizorn10, Gu10} including algorithms based on the
Multi-scale Entanglement Renormalization Ansatz\cite{MERA} (MERA) and Projected Entangled-Pair States.\cite{PEPS2, PEPS1, Jordan08, morePEPS, morePEPS2, Orus09} The fundamental ingredient, common to all approaches, is to incorporate fermionic statistics directly into the tensor network by regarding the tensors as linear maps of anticommuting degrees of freedom, an idea recently generalized to anyonic statistics.\cite{Koenig10,Pfeifer10} The main goal of fermionic tensor network methods is to address strongly correlated fermionic models, which suffer from the negative sign problem in Quantum Monte Carlo.\cite{sign}

Fermionic PEPS were first proposed in Ref.~\onlinecite{Kraus09} and have been discussed in several other papers.\cite{Barthel09, Shi09, Corboz10b, Pizorn10, Gu10} In Ref.~\onlinecite{Corboz10b} we provided a detailed account of how to adapt existing bosonic PEPS algorithms to the fermionic case, and we used the fermionic version of the infinite PEPS algorithms\cite{Jordan08,Orus09} to obtain benchmark results for three models with nearest-neighbor Hamiltonian in an infinite square lattice: ($i$) a model of free spinless fermions with a pairing potential, ($ii$) a model of interacting spinless fermions with a nearest-neighbor repulsion, and ($iii$) the well-known $t$-$J$ model. In the first case, a comparison of the numerical results with the exact solution showed that fermionic PEPS could reproduce ground-state energies and short-range correlators satisfactorily, with a larger degree of accuracy in gapped phases than in gapless ones. For the model of interacting spinless fermions, PEPS yielded significantly lower variational energies than obtained by mean-field theory (restricted Hartree-Fock theory), which enabled to determine the phase diagram more accurately. For the $t$-$J$ model, PEPS energies are comparable (or even better in some cases) than variational Monte Carlo based on Gutzwiller-projected ansatz wave functions. 

As with any new approach, systematic benchmarking of fermionic PEPS algorithms is important in order to establish their range of applicability. The results of Ref.~\onlinecite{Corboz10b}, while limited to three specific models, were a first step in this direction. A key question to be addressed is how good a fermionic PEPS is in practice, as a variational ansatz, at approximately representing the ground state of fermionic models. Of course, the precise answer to this question will depend on the specific model under consideration. However, insight on how fermionic PEPS methods generally perform in certain circumstances, e.g. in a given gapped phase, may be obtained from models where an exact solution or previous numerical results by other methods are already available. Such insight is essential in order to subsequently assess the validity of fermionic PEPS results obtained in more relevant (and challenging) scenarios, such as in exploring the ground state phase diagram of the $t$-$J$ model, which was addressed in Ref.~\onlinecite{Shi09}, or of the Hubbard model. 

Thus, one of the main goals of this paper is to further benchmark the performance of fermionic PEPS algorithms, by considering more complex models than those addressed in Ref.~\onlinecite{Corboz10b}. Specifically, here we will consider the effect of adding nearest-neighbor hopping terms to the three models of Ref.~\onlinecite{Corboz10b}. Recall that in many cases of interest it is desirable to consider a model where fermion particles can hop between nearest-neighbor sites (with amplitude $t$) as well as next-nearest neighbor sites (with amplitude $t'$).
For example, in effective models of high-T$_c$ superconductors (cuprates), it is estimated that the ratio $|t'/t|$ is of the order of $0.1-0.3$.\cite{Zhang88, Pavarini01}
A finite $t'$ can have several important effects on the system. For instance band-structure calculations\cite{Raimondi96, Pavarini01} and experimental analysis\cite{Tanaka04} suggest that the highest $T_c$ strongly depends on $t'/t$. 
Previous studies of the hole-doped $t$-$t'$-$J$ model revealed that a finite $t'<0$ can suppress magnetic order \cite{Tohyama94, Parcollet04, Spanu08} and enhance or suppress pairing correlations\cite{Shih04, Spanu08} (depending on the doping). It was also shown that $t'$ influences the formation of stripes.\cite{White99, Himeda02}  

In order to explore how well fermionic PEPS can reproduce such ground states, we need to extend the fermionic PEPS algorithm of Ref.~\onlinecite{Corboz10b}, based on simulating imaginary time evolution, to the case where the evolution is generated by a Hamiltonian that also contains next-nearest neighbor terms. Thus a second main goal of this paper is to explain how this is accomplished. The simulation of frustrated spin models with next-nearest neighbor terms with PEPS by imaginary time evolution was considered in Ref.~\onlinecite{Murg09}. In contrast to Ref.~\onlinecite{Murg09}, here we will explain how to generalize the so-called \textit{simple update} scheme for time evolution to the case of next-nearest neighbor terms (in addition to accounting for the fermionic character of the PEPS).

The paper is organized as follows: 
In Sec.~\ref{sec:method} we first explain how to update the PEPS during an imaginary time evolution in the presence of next-nearest neighbor Hamiltonian terms.
Then we discuss how to evaluate the expectation value of a next-nearest neighbor operator, as required e.g. in order to compute the energy of a PEPS in the case of nearest neighbor hopping $t'$. 
In Sec.~\ref{sec:results} we present a series of benchmark results for extensions, containing next-nearest neighbor hopping terms, of the three models addressed previously in Ref.~\onlinecite{Corboz10b}, namely ($i$) an exactly solvable model of free spinless fermions with $t'\neq 0$, ($ii$) the $t$-$t'$-$V$ model,\cite{Woul10} and ($iii$) the $t$-$t'$-$J$ model.\cite{Spanu08}
The accuracy and/or apparent convergence (as a function of bond dimension $D$) of ground state properties in these models are comparable to those previously obtained for the case $t'=0$.
Finally, Sec.~\ref{sec:conclusion} summarizes our findings and conclusions. 
For completeness, Appendix \ref{app:ctm} provides details on the two different corner-transfer-matrix (CTM) schemes used in order to evaluate expectation values from an infinite PEPS.

\section{Method} 
\label{sec:method}

As in Ref.~\onlinecite{Jordan08} for spin systems, we use an infinite PEPS with bond dimension $D$ to approximate the ground state of a Hamiltonian defined on an infinite square lattice, by simulating an evolution in imaginary time starting from some (random) initial state. 

The evolution itself is first approximated, through a Trotter-Suzuki decomposition, by a sequence of two-site gates.\cite{Jordan08} After applying each two-site gate, the affected bond in the PEPS has to be truncated, so that the evolved state is again represented by a PEPS with the same bond dimension $D$. This truncation implies choosing a $D$-dimensional subspace in the vector space associated to the bond index. In Ref.~\onlinecite{Corboz10b} we distinguished between two different truncation or update schemes. The first consists of choosing the subspace that best supports the wave function. This requires taking the whole PEPS wave-function into account during the update, i.e. the \textit{environment} has to be computed at every step in the imaginary time evolution.\cite{Jordan08} Alternatively, following Refs.~\onlinecite{morePEPS2}, one can update the PEPS as in the time-evolving block decimation (TEBD) method in one dimension\cite{TEBD} by means of a singular value decomposition (SVD). In this second, simpler option, referred to as \textit{simple update}, instead of considering the full environment, only local weights attached to the bonds of the PEPS are taken into account. In one dimensional systems with open boundary conditions, this choice of update is optimal, since the full environment can be encoded in the local weights. In two dimensional systems, however, the simple update is no longer optimal (a better chose of truncated space can be obtained by considering the whole environment) but it has a significantly lower computation cost as a function of bond dimension $D$, which allows to consider larger bond dimensions (in a suboptimal way) and potentially obtain more accurate results with the same computational cost. 
 
Here we will use the simple update. In this section we first discuss the additional steps needed to apply the simple update of Ref.~\onlinecite{Corboz10b} in the case of a two-site gate acting on next-nearest neighbor sites. Then we will also explain how to evaluate the expectation value of a next-nearest neighbor operator. 

\begin{figure}[htb]
\begin{center}
\includegraphics[width=8cm]{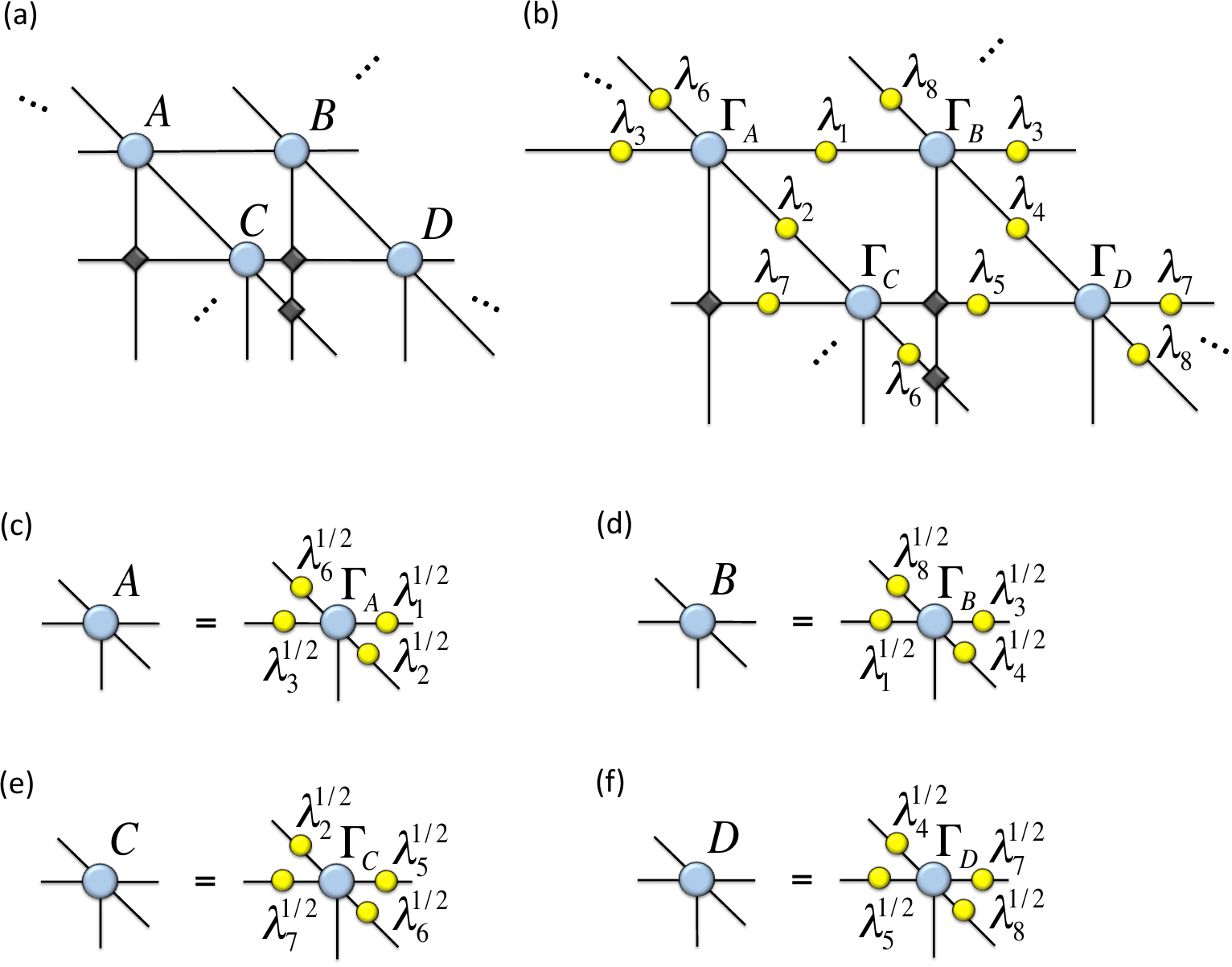}
\caption{(Color online) (a) Infinite PEPS with a four-site unit cell with tensors $A$, $B$, $C$, and $D$. (b) Representation of the same PEPS with diagonal matrices $\lambda_k$ on the bonds, which is used for the simple update. (c)-(f) Relation between the tensors in (a) and (b).}
\label{fig:structure}
\end{center}
\end{figure}

\subsection{Simple update for next-nearest neighbor terms}
\label{sec:simpleupdate}

\begin{figure}[htb]
\begin{center}
\includegraphics[width=7cm]{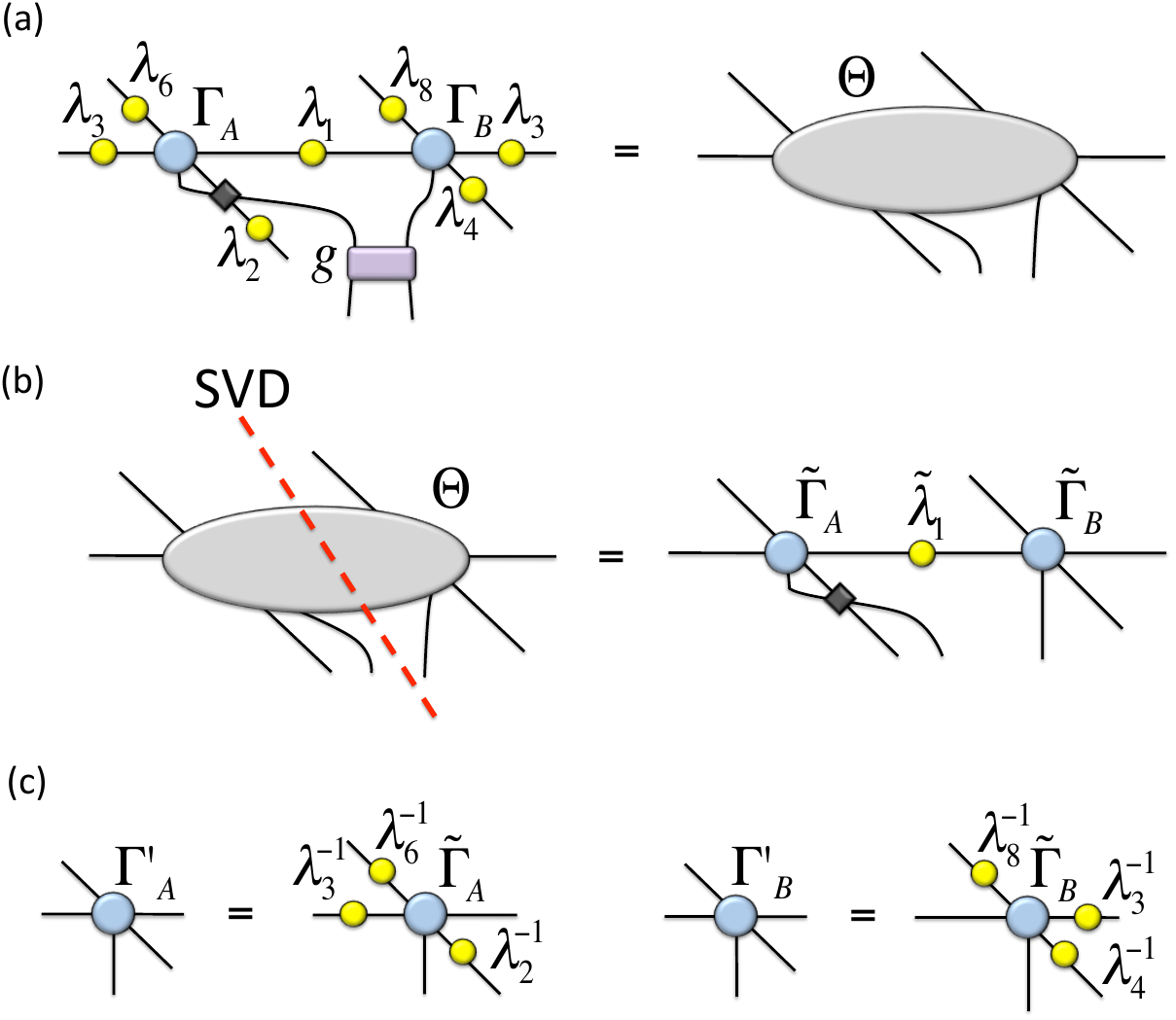}
\caption{(Color online) Simple update of a nearest-neighbor link. (a) A gate $g$ is applied to a link between tensors $\Gamma_A$ and $\Gamma_B$. Contracting the diagram, including all adjacent matrices $\lambda_k$, yields tensor $\Theta$. (b) Singular value decomposition of tensor $\Theta$ leads to tensors $\tilde \Gamma_A$, $\tilde \Gamma_B$ and diagonal matrix $\tilde \lambda_1$. (c)~The updated tensors $\Gamma'_A$, $\Gamma'_B$ are obtained by multiplying the corresponding unaltered inverse diagonal matrices $\lambda^{-1}_k$ to $\tilde \Gamma_A$ and $\tilde \Gamma_B$ as shown. The new diagonal matrix $\lambda'_1$ corresponds to the $D$ largest diagonal entries (singular values) of $\tilde \lambda_1$.}
\label{fig:nnupdate}
\end{center}
\end{figure}

In order to perform the simple update the fermionic PEPS in Fig.~\ref{fig:structure}(a) is recast into the form shown in  Fig.~\ref{fig:structure}(b), where the diagonal matrices $\lambda_k$ live on the bonds and the tensors $\Gamma_q$ live on the sites of the network. As already explained in Appendix B in Ref.~\onlinecite{Corboz10b} the simplified update for nearest-neighbor links consists of the three steps summarized in Fig.~\ref{fig:nnupdate}. 

\begin{figure}[htb]
\begin{center}
\includegraphics[width=8.5cm]{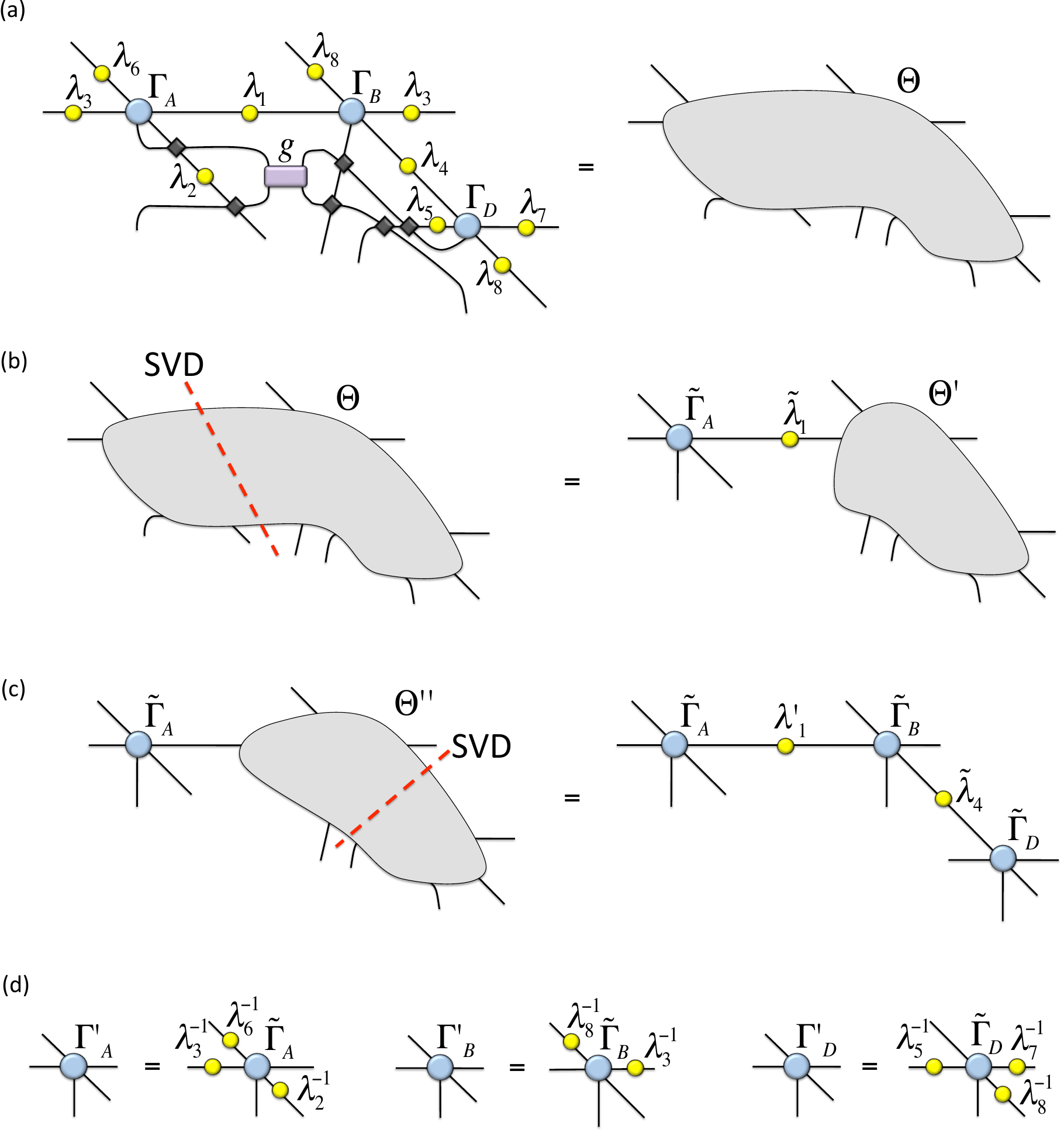}
\caption{(Color online) Simple update of a next-nearest neighbor link. 
(a)~A gate $g$ is applied to a link between tensors $\Gamma_A$ and $\Gamma_D$, via the tensor $\Gamma_B$. Contracting the diagram, including all adjacent matrices $\lambda_k$, yields tensor $\Theta$. (b)~Singular value decomposition of tensor $\Theta$ leads to tensors $\tilde \Gamma_A$, $\Theta'$, and  $\tilde \lambda_1$. (c) Tensor $\Theta''$ includes the weights $\lambda_1'$, obtained by keeping the $D$ largest diagonal entries of $\tilde \lambda_1$. Singular value decomposition of tensor $\Theta'$ leads to tensors $\tilde \Gamma_B$, $\tilde \Gamma_D$, and  $\tilde \lambda_4$.  (d)~The updated tensors $\Gamma'_A$, $\Gamma'_B$, and $\Gamma'_D$ are obtained from $\tilde \Gamma_A$, $\tilde \Gamma_B$, and $\tilde \Gamma_D$ as shown. The new diagonal matrix $\lambda'_4$ corresponds to the $D$ largest diagonal entries of $\tilde \lambda_4$.}
\label{fig:ADu}
\end{center}
\end{figure}

\begin{figure}[htb]
\begin{center}
\includegraphics[width=8.5cm]{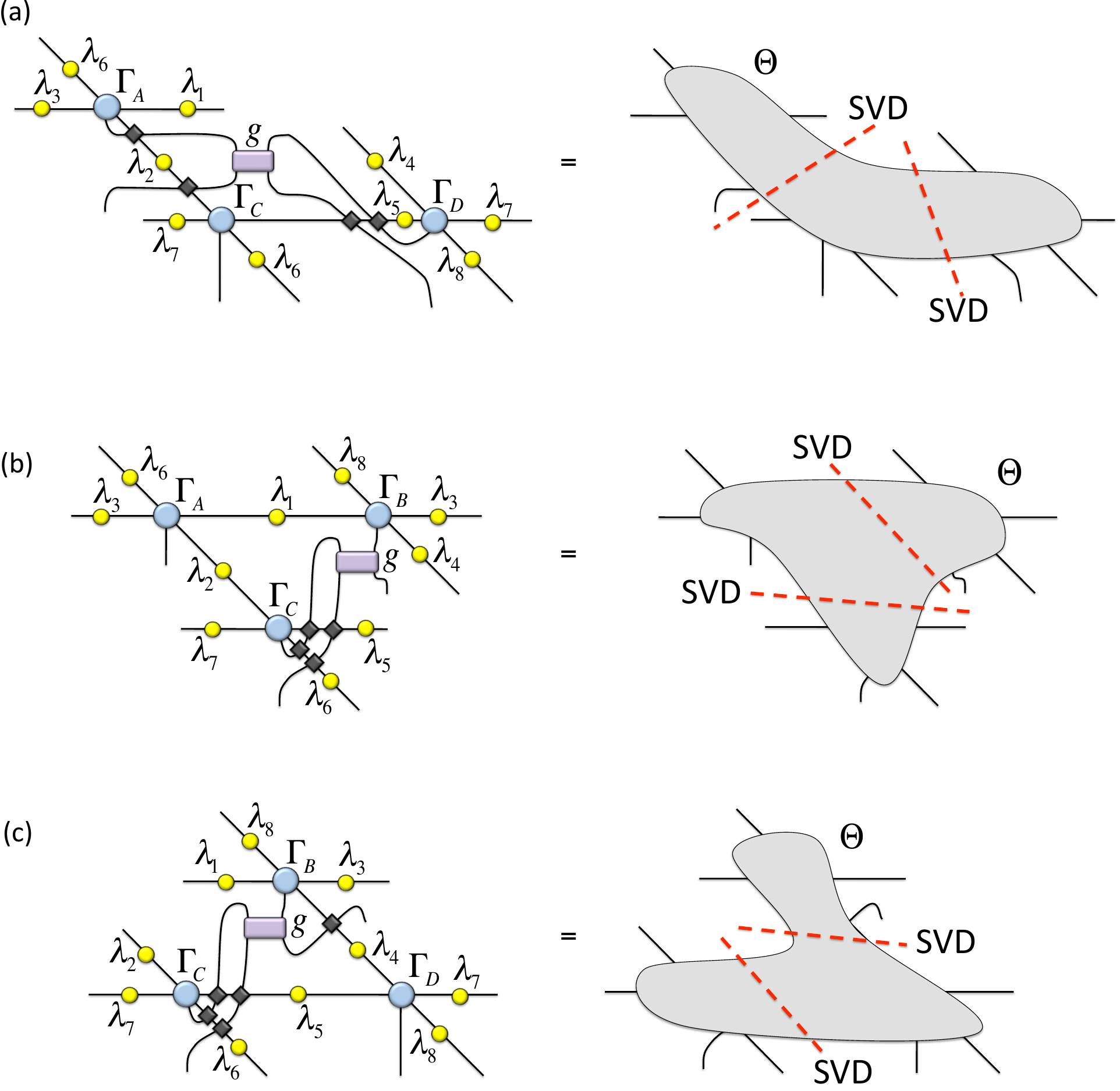}
\caption{(Color online) The remaining relevant diagrams of the next-nearest neighbor updates. The updated tensors are obtained in a similar way as explained in Fig.~\ref{fig:ADu}}.
\label{fig:otherupdates}
\end{center}
\end{figure}

The next-nearest neighbor update is performed in a very similar way, with the difference that three PEPS tensors and two diagonal matrices are updated at the same time, through two consecutive singular value decompositions. Figure~\ref{fig:ADu} illustrates the update for the link between tensors $\Gamma_A$ and $\Gamma_D$, via the tensor $\Gamma_B$ (including all adjacent diagonal matrices $\lambda_k$). Similarly one could also perform the update for the same link involving the tensor $\Gamma_C$ instead of $\Gamma_B$ as shown in Fig.~\ref{fig:otherupdates}(a). In practice we apply the square root of the gate to both combinations of tensors, in order to make the update more symmetric. In principle it is conceivable that also the order in which the singular value decompositions are made plays a role. However, in the cases studied we have not found a significant difference when changing the order. 

Figures~\ref{fig:otherupdates}(c)-(d) show the relevant diagrams for the update of the other diagonal link in the tensor network. As usual, crossings in the network have to be replaced by swap tensors (black diamonds) in order to account for the fermionic anticommutation rules. In total there are 8 different diagonal links for the $2\times 2$ unit cell, where each link is updated in two different ways (via two different intermediate tensors). 

We conclude this section with three remarks. Firstly, we note that the complexity of the update can be reduced by splitting off the parts of the tensors involved in the update by a singular value decomposition (see e.g. Fig.~32 in Ref.~\onlinecite{Corboz10b}). Secondly, the same update may of course be used also for bosonic and spin systems, with the simplification that crossings do not need to be taken into account, since bosonic and spin operators commute. Finally, note that an update for next-to-next nearest neighbor interaction can be implemented in a very similar way, since it also involves three PEPS tensors (arranged on a line). This case, not further considered in the paper, requires a larger unit cell of size $3\times 3$. 

\subsection{Computing expectation values of next-nearest neighbor terms}
\label{sec:expect}

\begin{figure}[htb]
\begin{center}
\includegraphics[width=8.5cm]{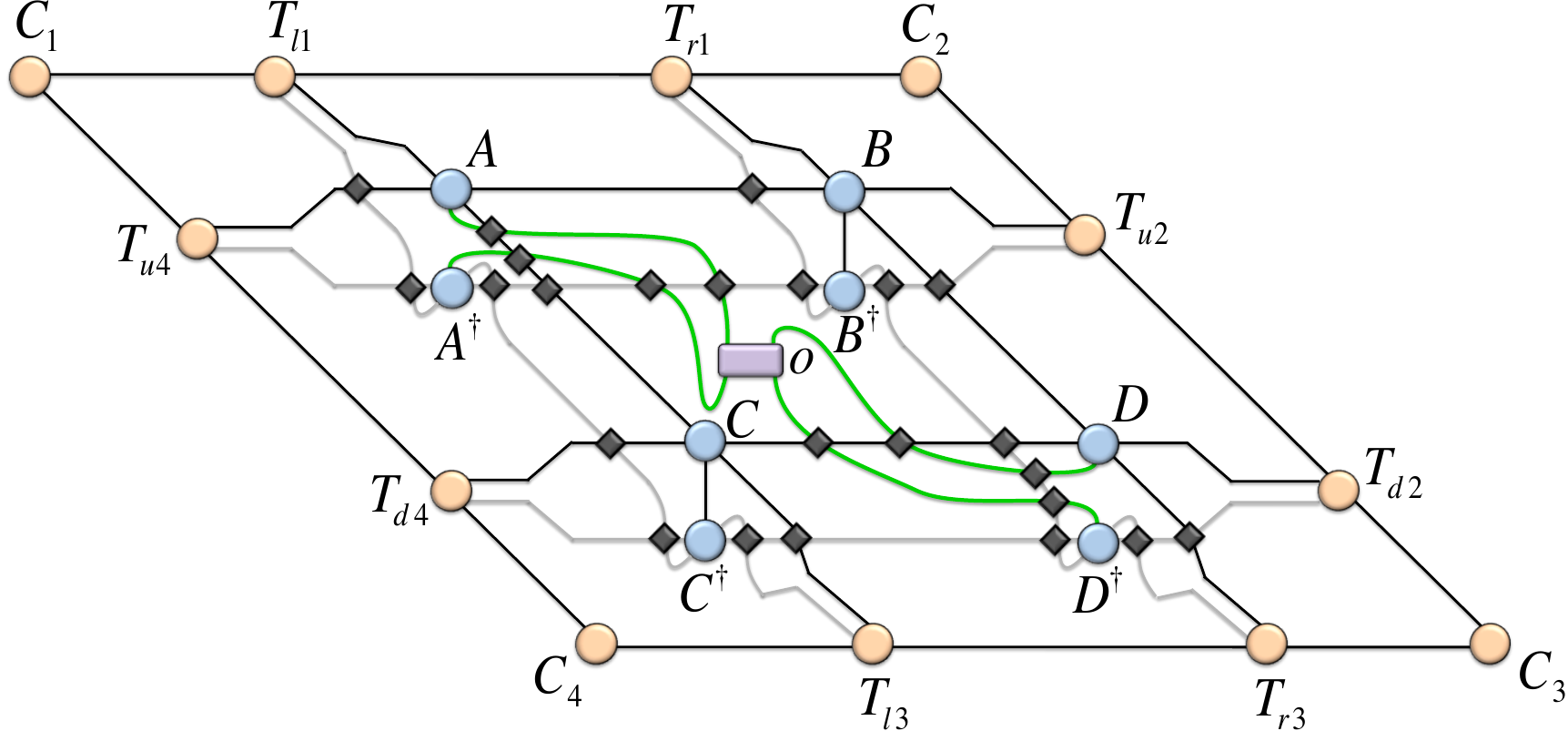}
\caption{(Color online) Tensor network to compute the expectation value of next-nearest neighbor operator $o$ acting between tensors $A$ and $D$.}.
\label{fig:ADham}
\end{center}
\end{figure}

In order to compute the expectation value of a next-nearest neighbor operator, one proceeds in a similar way as explained in Sec.~III B of Ref.~\onlinecite{Corboz10b} for nearest neighbor operators. First, one has to compute the environment ${\cal E}^{[{AB\atop{CD}}]}$ for the tensors $A, B, C, D$, which accounts for the infinite lattice surrounding the $2\times2$ unit cell formed by these tensors.
In Ref.~\onlinecite{Corboz10b} this was done with the directional corner transfer matrix (CTM) method.\cite{Orus09} Besides this scheme, in the present work we also use another variant of the CTM approach, as discussed in Appendix~\ref{app:ctm}.
The CTM algorithm yields the four corner tensors $C_1, C_2, C_3, C_4$ and the eight edge tensors $T_{l1},  T_{r1},  T_{u2},  T_{d2}, T_{l3},  T_{r3},  T_{u4},  T_{d4}$ shown in Fig.~\ref{fig:ADham}, which altogether constitute the environment ${\cal E}^{[{AB\atop{CD}}]}$. 

Next, one connects the four tensors $A, B, C, D$ together with their complex conjugates to the environment, and joins the physical legs to the operator accordingly, as exemplified in Fig.~\ref{fig:ADham} for a next-nearest neighbor operator acting between $A$ and $D$. 
Since the wave function encoded by the iPEPS is not normalized, the value obtained by contracting the tensor network in Fig.~\ref{fig:ADham} has to be divided by the norm of the iPEPS, which is simply obtained by replacing the two-site operator $o$ by the identity operator in Fig.~\ref{fig:ADham}. 

Evaluating a two-site operator $o$ linking tensors $B$, $C$ can be done in a similar way, simply by reconnecting the legs of the operator $o$ (highlighted in green in Fig.~\ref{fig:ADham}) to the physical legs of $B$ and $C$ accordingly. Finally, the expectation value of the remaining six next-nearest terms can be obtained analogously by first generating the environments for the plaquettes $[{BA\atop{DC}}], [{DC\atop{BA}}], [{CD\atop{AB}}]$.

\section{Benchmark results.}
\label{sec:results}
In this section we provide benchmark results for three fermionic models on an infinite square lattice with nearest neighbor and next-nearest neighbor hopping terms, with amplitudes $t$ and $t'$. Each model is an extension of an analogous model with only nearest neighbor hopping term, that is with $t'=0$, previously addressed in Ref.~\onlinecite{Corboz10b}.

As initial condition, the tensors of the infinite PEPS are chosen randomly. In some cases we observe that the state resulting from the time evolution depends on the initial condition (seed), which might be due to the local character of the simple update (cf. Sec.~\ref{sec:simpleupdate}). We therefore typically run several (of the order of 10) simulations with different seeds, and pick the state with lowest energy. In some cases a good choice is to initialize a PEPS from a previously converged PEPS with smaller $D$, but this does not always lead to the state with lowest energy. A converged PEPS for a certain Hamiltonian usually provides a good initial condition for simulations with slightly different Hamiltonian parameters, provided both states are in the same phase.

Evaluating the expectation value of observables requires computing an environment, which in turn requires introducing a second bond dimension $\chi$ associated to additional truncations. \cite{Jordan08,Orus09,Corboz10b} In all present simulations the bond dimension $\chi$ of the environment has been chosen to be sufficiently large so that the expectation values of local observables do not significantly change when further increasing $\chi$. Typical values are $\chi=36$ for $D=2$, $\chi=48$ for $D=4$, and $\chi=64$ for $D=6$ and $D=8$.  

\subsection{Free fermions including a pairing potential}
\label{sec:sf}
We first consider a model of free spinless fermions given by the Hamiltonian
\begin{eqnarray}
	H =&& t \sum_{\langle ij \rangle} [\hc_i^{\dagger}\hc_j + H.c.]  
	- \gamma \sum_{\langle ij \rangle}  [\hc_{i}^{\dagger}\hc_{j}^{\dagger} + H.c.]  \nonumber\\
	 &&  + \,\,\, t' \sum_{\langle \langle ij \rangle \rangle} [\hc_i^{\dagger}\hc_j + H.c.] 
	 - 2\lambda \sum_i \hc_{i}^{\dagger} \hc_i   ,
\label{eq:free}
\end{eqnarray}
with $\langle ij \rangle$ and $\langle \langle ij \rangle \rangle$ denoting the sum over nearest and next-nearest neighbor pairs, respectively. 
In the following we fix the hopping amplitude $t$ to $1$ and the pairing potential $\gamma$ to $1$, and consider a chemical potential $\lambda$ in the range $[1,4]$. For $t'=0$ the model reduces to the one studied in Sec.~IV of Ref.~\onlinecite{Corboz10b}, and is gapped for $\lambda>2$ and critical for $\lambda \le2$. \cite{Li06} For $t'\neq 0$, the location of the transition $\lambda_C$ between gapped and gapless phases depends on~$t'$. 

\begin{figure}[htb]
\begin{center}
\includegraphics[width=8cm]{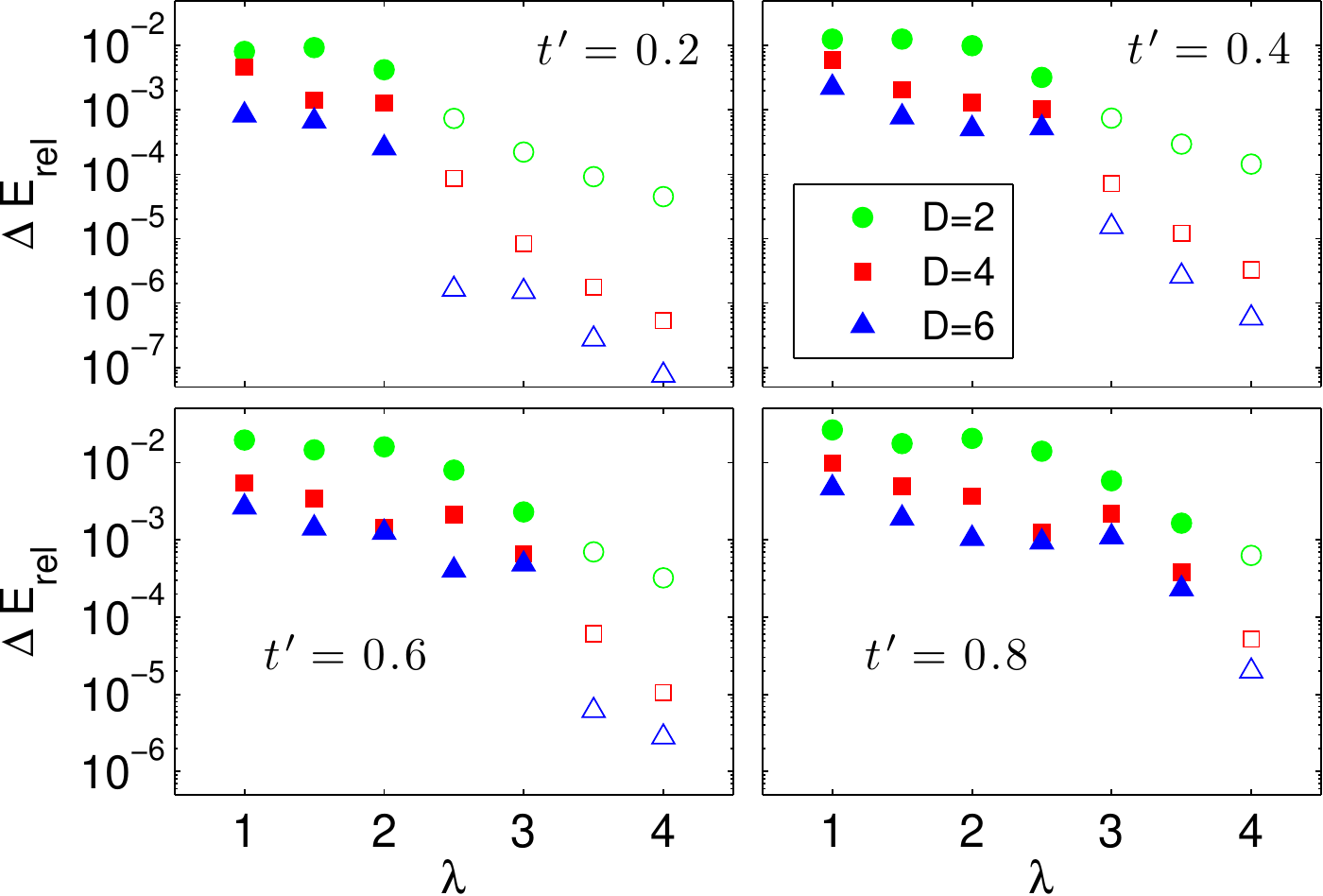}
\caption{(Color online) Relative error of the ground state energy of the free spinless fermion model \eqref{eq:free} as a function of $\lambda$, for different values of $D$ and $t'$. Full and open symbols correspond to critical and gapped phases, respectively.}
\label{fig:ttp}
\end{center}
\end{figure}

Figure \ref{fig:ttp} shows the relative error of the ground state energy as a function of $\lambda$ for different $D$ and $t'$. As in the case for $t'=0$ the energies are improved upon increasing $D$, and the accuracy is higher in the gapped phase (open symbols) than in the critical phase (full symbols). In general the error is larger than in the $t'=0$ case (cf. Ref.~\onlinecite{Corboz10b}), but still of the order of $10^{-3}$ ($10^{-5}$) in the critical (gapped) phase for $D=6$.

\begin{figure}[htb]
\begin{center}
\includegraphics[width=8cm]{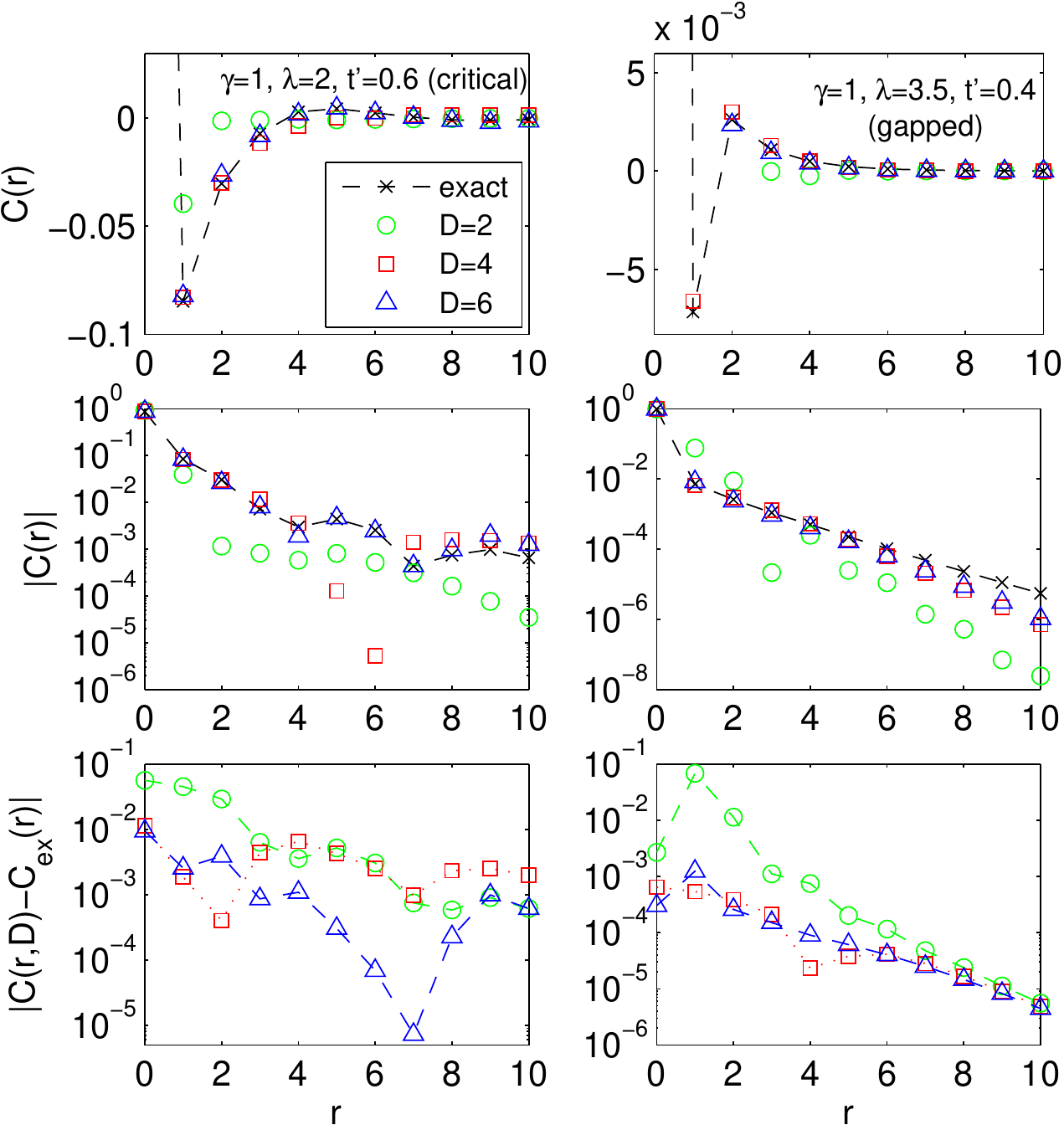}
\caption{(Color online) Upper panels: Correlation function $C(r)=\langle \hc^\dagger_i \hc_{i+r} \rangle$ as a function of distance r (in x-direction) in the gapless (left) and gapped (right) phase of the free-fermion model  \eqref{eq:free}. Middle panels: Absolute value of $C(r)$ in semi-logarithmic scale. Lower panels: The difference between the simulation result $C(r,D)$ and the exact result $C_{ex}(r)$ for different values of $D$. }
\label{fig:ttp_corrs}
\end{center}
\end{figure}

Figure~\ref{fig:ttp_corrs} shows the two-point correlation function
\begin{equation}
	C(r) \equiv \langle \hc^\dagger_i \hc_{i+r} \rangle,
	\label{eq:CdagC}
\end{equation}
as a function of distance $r$ between the two sites (in x-direction). The numerical results are seen to approach the exact values with increasing $D$, with correlations at short distances being better reproduced than correlations at long distances (see middle panels). Also in this case, the accuracy is better in the gapped phase ($\gamma=1, \lambda=3.5, t'=0.4$) than in the critical phase ($\gamma=1, \lambda=2, t'=0.6$).

\subsection{The $t$-$t'$-$V$ model}
Next we study the $t$-$t'$-$V$ spinless fermion model and compare the infinite PEPS results with the mean-field studies from Ref.~\onlinecite{Woul10}, based on Hartree-Fock (HF) theory restricted to states invariant under translations by two sites. The Hamiltonian reads
\begin{eqnarray}
	H =  &-& t \sum_{\langle ij \rangle} [\hc_i^{\dagger}\hc_j + H.c.]  + V \sum_{\langle ij \rangle}  \hc_i^{\dagger}\hc_i \hc_j^{\dagger}\hc_j \nonumber \\
	 &-& t' \sum_{\langle \langle ij \rangle \rangle} [\hc_i^{\dagger}\hc_j + H.c.] - \mu \sum_i \hc_{i}^{\dagger} \hc_i,
\label{eq:ttV}
\end{eqnarray}
with $V$ being the nearest-neighbor interaction strength, and $\mu$ the chemical potential. As an example we study the transition between a metal phase at low electron density $n$ to a charge-density-wave (CDW) phase at half filling ($n=0.5$), for fixed parameters $t=1$, $t'=-0.4$, and $V=2$. [A similar study was done in Ref.~\onlinecite{Corboz10b} in the case of $t'=0$.]

\begin{figure}[htb]
\begin{center}
\includegraphics[width=8cm]{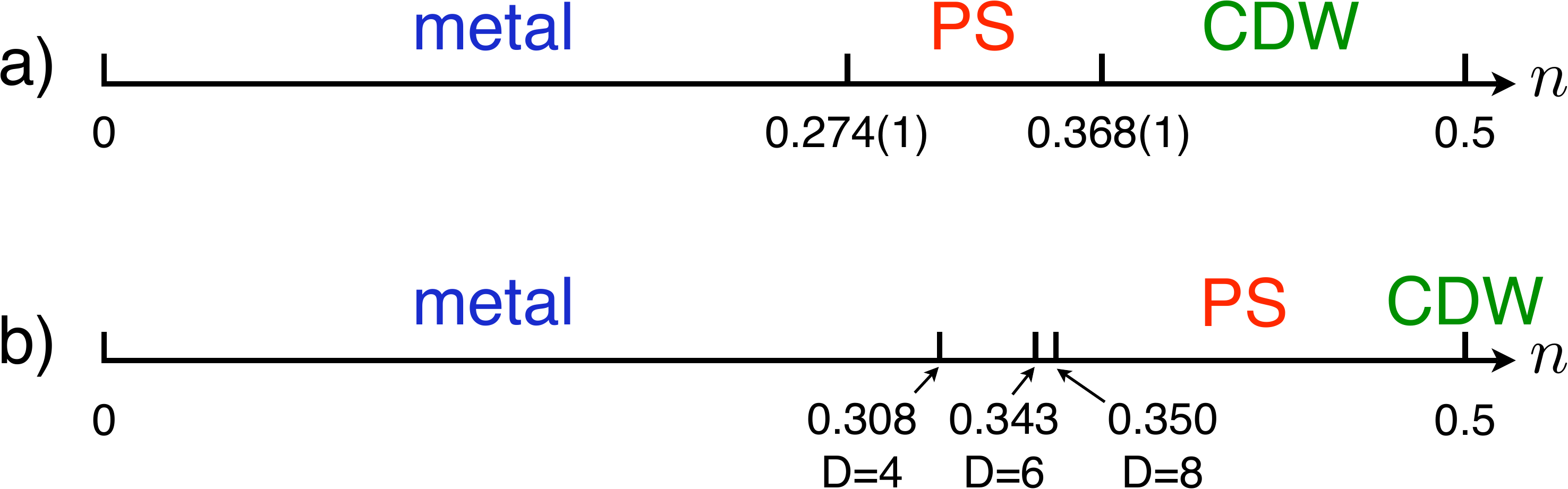}
\caption{(Color online) (a) Mean-field phase diagram for fixed parameters $V=2$ and $t'=-0.4$ as a function of particle density $n$, obtained by Hartree-Fock (HF) restricted to states invariant under translations by two sites.\cite{Woul10} The charge-density-wave (CDW) phase is separated from the metal by a region of phase separation (PS). The CDW phase is stable upon doping. b) Phase diagram obtained with iPEPS for different bond dimensions $D$ for the same parameters as in a). A stable CDW phase is only found at half filling. The phase boundary to the PS region is shifted towards higher values of $n$ with increasing $D$.}
\label{fig:ttV_pd}
\end{center}
\end{figure}

The HF study predicts a first order phase transition between the metallic phase for densities $0 <n\le 0.274(1)$ and a CDW phase, which is thermodynamically stable upon doping in the range $0.368(1)\le n \le 0.5$, as illustrated in Fig.~\ref{fig:ttV_pd}(a). The region in $0.274(1)<n<0.368(1)$ is unstable, corresponding to phase separation (PS) between the two states. The study was done for systems of size $100 \times 100$. 

\begin{figure}[htb]
\begin{center}
\includegraphics[width=8cm]{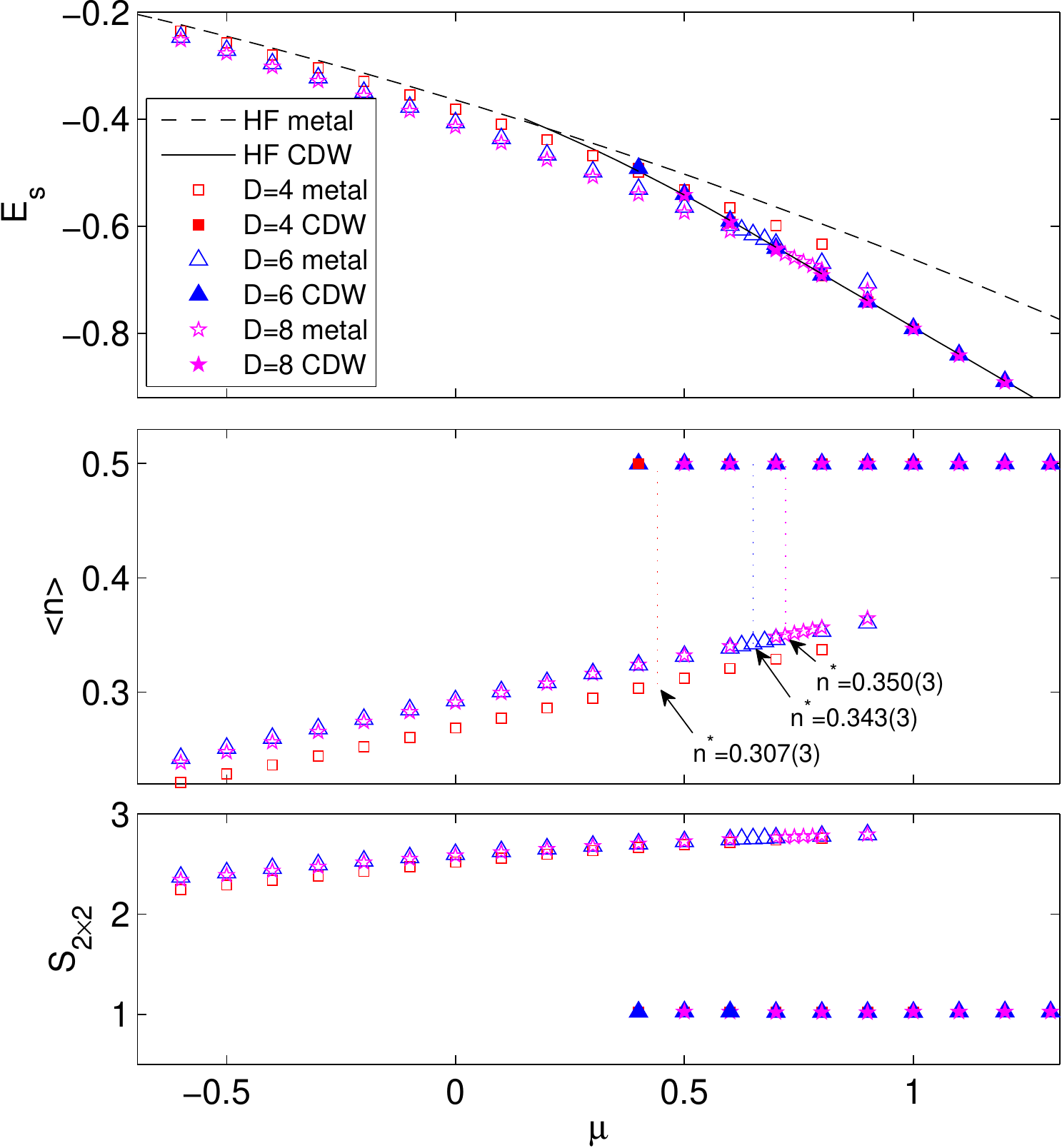}
\caption{(Color online) Upper panel: Energy per site of the $t$-$t'$-$V$ model \eqref{eq:ttV} as a function of chemical potential $\mu$ for $V=2$ and $t'=-0.4$, obtained with (restricted) Hartree-Fock (HF)\cite{Woul10} and infinite PEPS. The first order phase transition between the metal phase and the charge-density wave (CDW) phase occurs at a value $\mu^*$ where the two corresponding energies cross. 
Middle panel: Particle density $n$ as a function of chemical potential in the two phases. 
At the first order phase transition point $\mu^*$, $n$ jumps from a certain value $n^*$ in the metal phase to $n=0.5$ in the CDW phase. For densities in between $n^*$ and $n=0.5$ the system exhibits phase separation. In contrast to the HF study, infinite PEPS simulations do not yield a stable doped CDW phase for $D\ge 4$. 
Lower panel: Entanglement entropy of a $2\times2$ block in the two phases, illustrating that there is substantially more entanglement in the metal phase than in the CDW phase.}
\label{fig:ttV}
\end{center}
\end{figure}

Figure~\ref{fig:ttV} shows a comparison of the energies in the two phases, obtained with HF and infinite PEPS. As explained in Ref.~\onlinecite{Corboz10b} for the case $t'=0$, the crossing of the two energies is obtained by starting from a state deep in the metal (CDW) phase and then increase (decrease) $\mu$ across the transition. Similarly as for $t'=0$, PEPS energies in the gapped CDW phase (for $n=0.5$) do not differ significantly from the HF energies.
However, with increasing $D=4,6,8$, PEPS energies do differ significantly from HF energies in the metal phase close to the transition. This produces a shift of transition point $\mu^*$ to larger values of $\mu$, corresponding to a larger value of the density $n$, as shown in Fig.~\ref{fig:ttV_pd}(b). Also, in contrast with the HF prediction, which predicts a stable doped CDW phase, we do not find a stable doped CDW phase for $D \ge 4$, 
i.e states in the doped CDW phase obtained by our PEPS computation exhibit a higher energy than states in the metal phase or in the CDW phase at half filling. Thus, we only find a stable CDW phase at exactly half filling (for the model parameters under consideration).

The amount of entanglement in the gapped CDW phase is relatively low, as can be seen in the lower panel in Fig.~\ref{fig:ttV}, showing the entanglement entropy of a $2\times2$ block in the system. Therefore, a PEPS with small bond dimension is already sufficient in order to obtain an accurate description of the ground state. In the metal phase, however, the entanglement entropy is considerably higher. The energies have not yet converged as a function of the bond dimension $D$ for $D=8$, and thus further corrections to the energy can be expected for larger values of $D$. This phase appears to be particularly difficult to represent by the PEPS. This is not surprising. In the presence of a 1D Fermi surface the entanglement entropy exhibits a logarithmic multiplicative correction to the area law, as shown in the case of free fermions.\cite{Li06,LOGCORR} A PEPS representation, however, can only reproduce an strict area law of the entanglement entropy, \cite{PEPS2} and therefore it is unclear that the ansatz can offer an accurate approximation of the ground state of a metallic phase. Nevertheless, our results also show that, with increasing bond dimension $D$, the PEPS can still be used to obtain a systematic improvement over HF results. 
Since the phase boundary does not change much when comparing the $D=6$ with the $D=8$ simulations, it seems likely that the $D=8$ result is already close to the exact one. 

We conclude this section with two remarks. Firstly, we observed states in the metal phase close to the transition that exhibit a slight density modulation between sublattices $A$ and $B$ in the lattice, similarly as in the CDW phase. However, this modulation becomes weaker with increasing $D$, which strongly suggests that this symmetry breaking of translation invariance is a numerical artifact due to small $D$ rather than a real physical feature. 

Secondly, it is conceivable that the $t$-$t'$-$V$ model exhibits also CDW phases other than the checkerboard-ordered phase at half filling, i.e. with a period larger than 2. 
%
Indeed, in some simulations we observed states where the density of e.g. sublattice $A$ oscillates as a function of the CTM steps, which could be an indication for a CDW phase with a period larger than 2. However, in order to represent such states accurately by a PEPS, either a rather large bond dimension or a larger unit cell than the ones employed in this work would be required. We therefore point out that the final phase diagram is likely to be different than the one presented in Fig~\ref{fig:ttV_pd}.

\subsection{$t$-$t'$-$J$ model}
Finally, we present benchmark results for the energy of the $t$-$t'$-$J$ model, given by the Hamiltonian
\begin{eqnarray}
H=\! &-& \! t \sum_{\langle ij \rangle \sigma} [ \tilde{c}_{i \sigma}^{\dagger}\tilde{c}_{j\sigma}  + H.c.] +  J\sum_{\langle ij \rangle}  ( \hat S_i \hat S_j - \frac{1}{4} \hat n_i \hat n_j) \\
 \!&-& \! t' \sum_{\langle \langle ij \rangle \rangle \sigma} [ \tilde{c}_{i \sigma}^{\dagger}\tilde{c}_{j\sigma}  + H.c.] -   \mu \sum_{i} \hat n_i 
 \end{eqnarray}
with $\sigma=\{\uparrow,\downarrow\}$ the spin index, $\hat n_i=\sum_\sigma \hc^\dagger_{i \sigma} \hc_{i \sigma}$ the electron density and $\hat S_i$ the spin $1/2$ operator on site $i$, and $\tilde{c}_{i\sigma}=\hc_{i\sigma} ( 1 - \hc^\dagger_{i \bar \sigma} \hc_{i \bar \sigma})$.

Here we compare our results of the energy with variational Monte Carlo (VMC) and fixed-node Monte Carlo (FNMC) results from Ref.~\onlinecite{Spanu08}, which are based on Gutzwiller-projected ansatz wave functions including spin and density Jastrow factors. 

\begin{figure}[htb]
\begin{center}
\includegraphics[width=7.8cm]{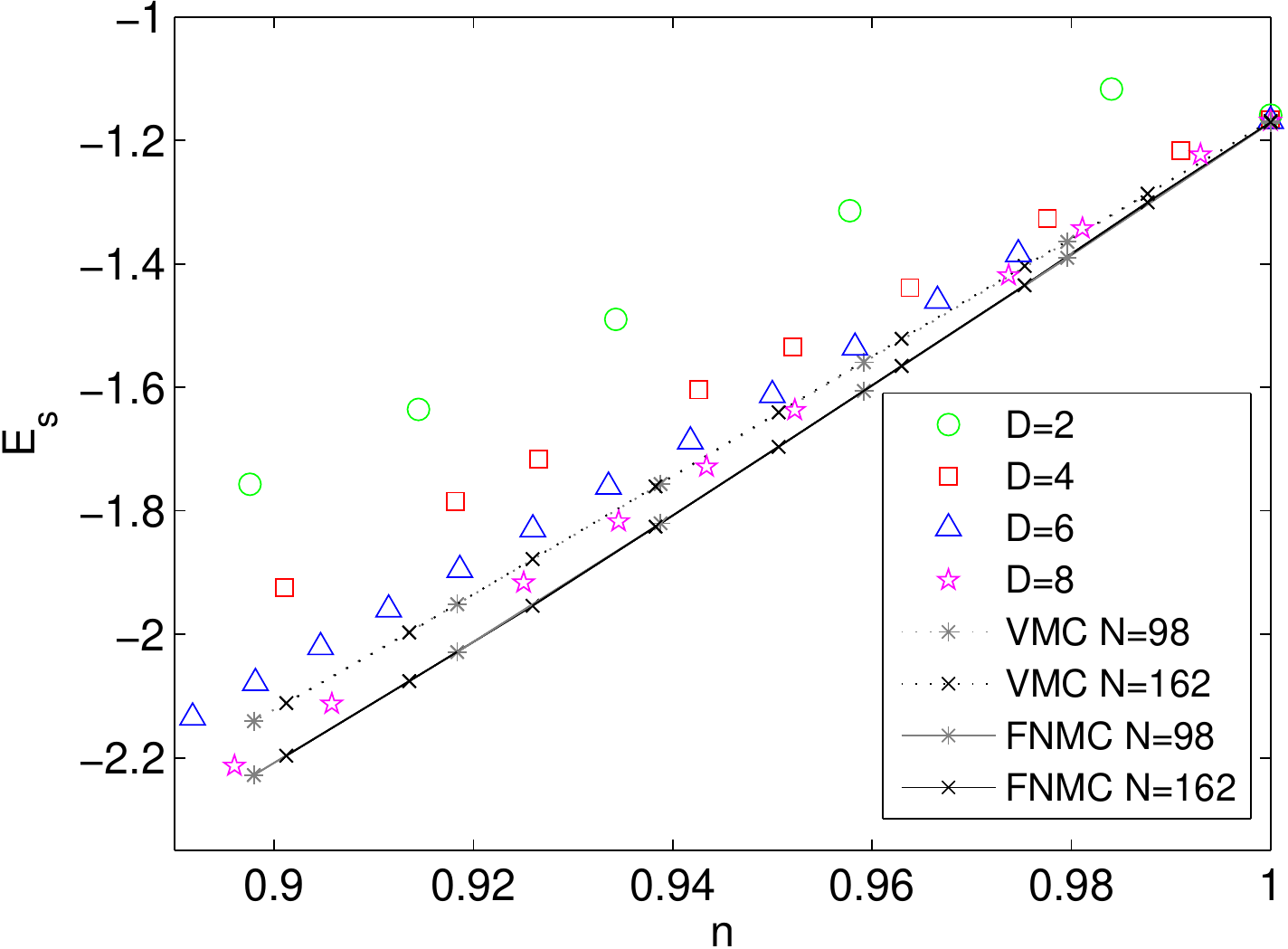}
\caption{(Color online) Energy per site (with the chemical potential term subtracted) as a function of particle density $n$ of the $t$-$t'$-$J$ model, with $J/t=0.4$ and $t'/t=-0.2$. The best iPEPS results for $D=8$ lie in between the VMC and the FNMC results.  }
 \label{fig:ttJ}
\end{center}
\end{figure}

Figure~\ref{fig:ttJ} shows the energy as a function of particle density $n$ for $J/t=0.4$ and $t'/t=-0.2$. Similarly as in the case $t'=0$ of Ref.~\onlinecite{Corboz10b}, the results for $D<8$ have a higher energy than VMC, but for $D=8$ the energies are comparable or even lower than VMC. Thus, the additional $t'$ does not seem to change the accuracy of the ansatz significantly (compared to Fig.~26 in Ref.~\onlinecite{Corboz10b}). 
The FNMC results, however, are still considerably lower than the $D=8$ results. Note that the Monte Carlo energies are for a finite lattice size (with 98 and 162 sites), and that the energy increases with increasing system size. Thus, the FNMC energy in the thermodynamic limit is slightly higher than shown in the plot. Finally we point out that for $D=8$ the maximal dimension $\chi$ we used is 64. It is conceivable that the energies still change slightly when increasing $\chi$ further, but we do not expect a significant change.

\section{Discussion and conclusion}
\label{sec:conclusion}
In this paper we have explained how to extend the fermionic PEPS algorithm for infinite lattices presented in Ref.~\onlinecite{Corboz10b} to models including next-nearest neighbor terms in the Hamiltonian. This opens the possibility to study a much larger variety of models, which are relevant e.g for high-temperature superconductivity.\cite{Zhang88}

The benchmark results in Sec.~\ref{sec:results}, involving three different models, indicate that the accuracy of the ground state energy is comparable to the case where only nearest-neighbor terms are present in the Hamiltonian. The present results are compatible with, and provide additional evidence in favor of, the conclusions of Ref.~\onlinecite{Corboz10b}, where it was indicated that a fermionic PEPS seems to offer a more accurate description of fermionic for gapped phases than for critical phases, among which two types (I and II) need to be distinguished. 

\textit{Gapped phases}.--Our simulations produced high accuracies (or better convergence as a function of bond dimension $D$) for ground state energies for gapped systems, such as the free spinless fermion model \eqref{eq:free} for large $\lambda$ and the interacting spinless fermion model \eqref{eq:ttV} at half filling. The corresponding ground states exhibit a comparatively small amount of entanglement and can therefore be approximated accurately by a PEPS with small bond dimension $D$. 

\textit{Gapless phases of type I}.--Gapless systems with a finite number of zero modes in their spectrum (that is, without a 1D Fermi surface) are in general more entangled than gapped systems, but they are believed to still obey the area law of the entanglement entropy, which a PEPS is known to be able to reproduce. \cite{PEPS2} Therefore, a PEPS is still expected to offer an accurate description of the ground state, provided certain bond dimension $D$, in general larger than in the gapped case, is used. This category of states includes, for the free spinless fermion model \eqref{eq:free}, the gapless p-wave paired phase corresponding to small $\lambda$; and, for the $t$-$t'$-$J$ model, both the antiferromagnetic phase at half-filling and the (expected) d-wave paired phase in the doped case. 
A remarkable achievement of fermionic PEPS simulations is that they yield better or comparable energies than the usual Gutzwiller projected ansatz wave functions for the doped $t$-$t'$-$J$ model for $D=8$. However, a larger bond dimension $D$ would be needed to attempt to match the energies of state-of-the-art fixed-node Monte Carlo (FNMC).\cite{Spanu08} 

We note that even if the energy in such gapless phases is obtained with a few digits of accuracy (of the order of $0.1\%$ for the model \eqref{eq:free} with $D=6$), it is still unclear whether the PEPS reproduces other relevant properties of the ground state accurately. Here we found that the correlation function \eqref{eq:CdagC} for the free fermion model \eqref{eq:free} is satisfactorily reproduced for short-range distances. However, there are other known cases, e.g. the Heisenberg antiferromagnet model, where the accuracy of the order parameter is two orders of magnitude worse than the accuracy of the energy for a $D=5$ PEPS on an infinite lattice.\cite{Bauer09} Therefore, the reliability of PEPS results must be carefully checked on a case by case basis. Nevertheless, and taking into account that there are no exact methods available to address systems of strongly correlated fermions, we believe that fermionic PEPS and, more generally, fermionic tensor networks, offer a useful approach to such systems that complements other approaches such as fixed-node Monte Carlo,\cite{FNMC} diagrammatic Monte Carlo,\cite{Prokofev98} cluster dynamical mean-field theory,\cite{Maier05} or Gaussian Monte Carlo.\cite{GQMC}
 
\textit{Gapless phases of type II}.--Gapless systems with a 1D Fermi surface, such as the metal phase of the interacting spinless fermion model \eqref{eq:ttV}, are known to display logarithmic multiplicative corrections to the area law of the entanglement entropy. This logarithmic violation can not be reproduced with a PEPS, and it is therefore unclear that fermionic PEPS methods will be able to accurately describe the ground state of such massively entangled phases. Nonetheless, our results for the metal phase of the interacting spinless fermion model \eqref{eq:ttV} show that, even in this case, PEPS with increasing values of the bond dimension $D$ can be used to obtain systematic improvements on mean-field energies, which in turn question the validity of the mean-field phase diagram. 

It might well be, however, that a proper characterization of the ground state of gapless phases with a 1D Fermi surface is simply beyond the reach of fermionic PEPS. In this case other techniques, such as one of the many methods which work particularly well in Fermi-liquid type phases at weak coupling (e.g. Refs.~\onlinecite{Prokofev98, GQMC}), or a specialized tensor network approach,\cite{Evenbly10} should be used instead.

We conclude by noticing that a larger bond dimension $D$, and therefore more accurate PEPS results, may be within reach in subsequent studies. This larger values of $D$ could be accessed e.g. by using more computer resources (possibly in a parallel architecture), by exploiting internal the symmetries of the the fermionic models\cite{Singh09} (e.g. particle conservation) and/or by employing Monte Carlo sampling techniques.\cite{Sandvik07} We also point out that with the same values of $D=2-8$ used in this paper, more accurate results may be obtained by employing the \textit{standard update} instead of the \textit{simple update} in the simulations, which, however, comes with a larger computational cost (cf. Ref.~\onlinecite{Corboz10b}).

{\it Acknowledgements.-} We thank J. de Woul, E. Langmann, and F. Becca for inspiring discussions and for providing their data from Refs.~\onlinecite{Woul10} and \onlinecite{Spanu08}. Valuable conversations with R.~Or\'us, B.~Bauer, and L.~Tagliacozzo, and support from the Australian Research Council (FF0668731, DP0878830, DP1092513) are acknowledged.

\appendix

\begin{figure}[htb]
\begin{center}
\includegraphics[width=8.5cm]{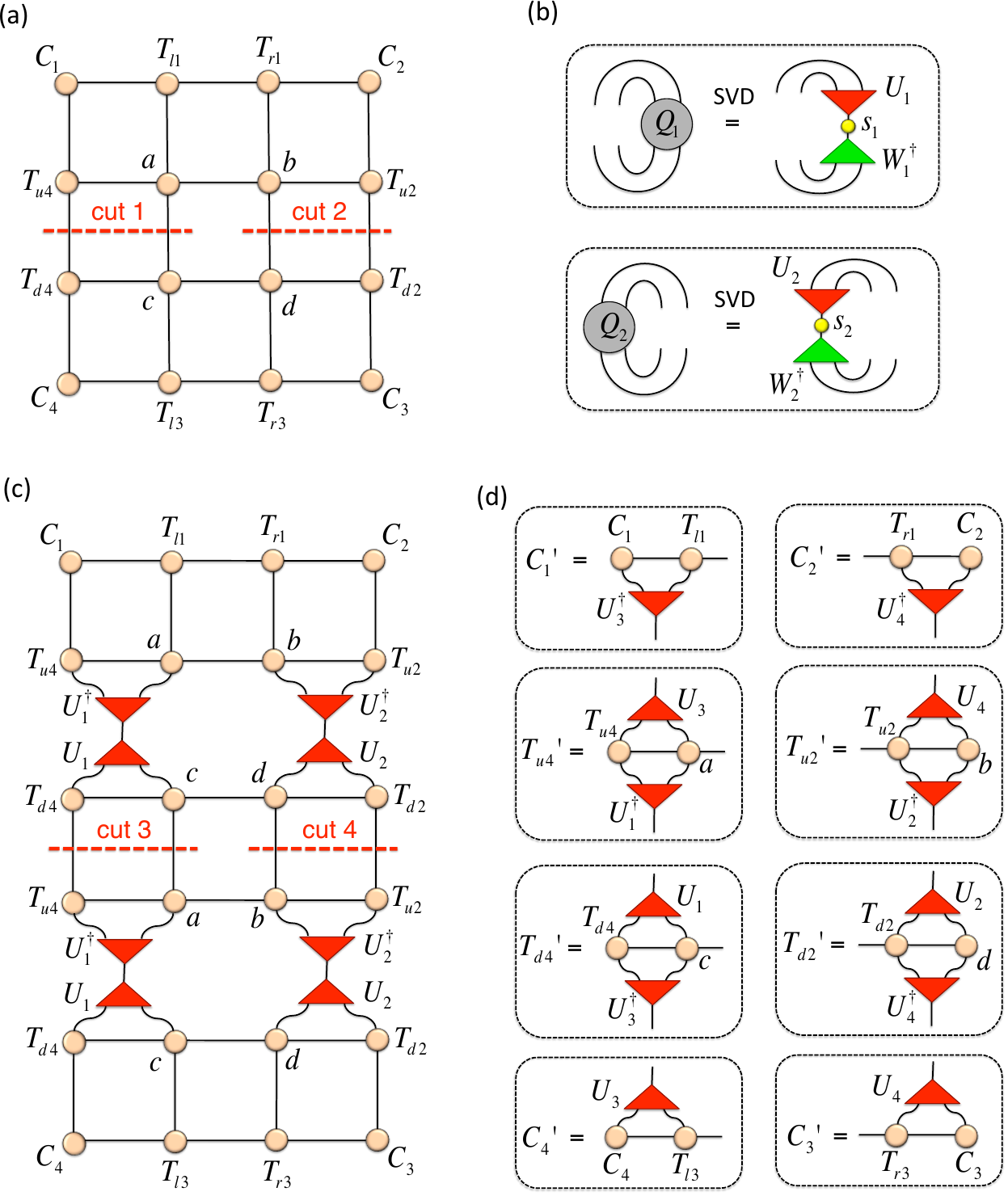}
\caption{(Color online) Left-right renormalization step of the anisotropic CTMRG method. (a) Tensor network of the reduced tensors $a$,$b$,$c$, and $d$ embedded in the environment. Cutting the lines as marked in the figure leads to tensors $Q_1$ and $Q_2$ shown in (b). (b) A singular value decomposition of the tensors $Q_1$ and $Q_2$ is performed, and only the $\chi$ largest singular values are kept. (c) Tensor network obtained by inserting two extra rows of tensors in the middle of the tensor network of (a), and by introducing an approximate resolution of the identity ${\mathbb I} \approx U_1^\dagger U_1$ in (each of the the two copies of) cut 1, and similarly for cut 2. [One could also use $W_1$ instead of $U_1$]. The cuts 3 and 4 define tensors $Q_3$ and $Q_4$, and a singular value decomposition of these tensors is performed, similarly as in (b), yielding tensors $U_3$ and $U_4$. (d)~Renormalized corner and edge tensors. Note that the top and bottom edge tensors have to be swapped after the update, i.e. $T_{l1}'= T_{r1}$, $T_{r1}'= T_{l1}$,  $T_{l3}'= T_{r3}$, $T_{r3}'= T_{l3}$.}
 \label{fig:NCTM}
\end{center}
\end{figure}

\section{Variant of the corner transfer matrix method}
\label{app:ctm}

The evaluation of the expectation value of a local observable from a PEPS requires the computation of the so-called \textit{environment},\cite{Jordan08,Orus09,Corboz10b} which accounts locally for the ground state wave function on the rest of the system, as explained in Sec.~\ref{sec:expect}. In an infinite system, corner transfer matrix (CTM) methods, originally introduced by Baxter,\cite{BaxterCTM} can be used to approximately compute this environment. In this case, the environment is approximated by four corner tensors $C_1$, $C_2$, $C_3$ and $C_4$ and eight edge tensors (or half-row transfer matrices) $T_{l1}$,  $T_{r1}$,  $T_{u2}$, $T_{d2}$, $T_{l3}$,  $T_{r3}$,  $T_{u4}$, and $T_{d4}$, as shown in Fig.~\ref{fig:ADham}. 

In the present work we applied two different CTM schemes: The first is the directional CTM method,\cite{Orus09} already used in our previous work;\cite{Corboz10b} the second scheme essentially corresponds to the CTM renormalization group (CTMRG) method from Ref.~\onlinecite{Nishino96} adapted to the anisotropic case and to a $2\times2$ unit cell, with subsequent coarse-graining moves in horizontal and vertical direction as in the directional CTM method. The main difference is that in the former scheme the four corners in the lattice are renormalized individually (see Refs.~\onlinecite{Orus09, Corboz10b} for details), whereas in the CTMRG approach the full environment is taken into account in each renormalization step. Figure~\ref{fig:NCTM} illustrates the left-right coarse-graining move, i.e. where the system size is increased by two lattice sites in the horizontal direction. In a similar way a top-bottom move is performed, which increases the system by two lattice sites in the vertical direction. These two moves are iterated until convergence is reached. 
Including more tensors in each renormalization step helps to better determine the relevant subspace to be kept during truncation, since more information about the system is taken into account.\footnote{A similar idea is pursued in the second renormalization group method.\cite{Xie09}} The disadvantage is that this scheme is more strongly limited by the machine precision of the computer. This can be understood by considering the isotropic case: In the directional CTM of Ref.~\onlinecite{Orus09} the spectrum of a single corner matrix $C$ is computed, whereas in the present scheme, which involves multiplying all four corner matrices, yields the fourth power of the same spectrum, with singular values (in this case, eigenvalues) expanding many more orders of magnitudes, and therefore more vulnerable to errors due to finite machine precision. 

We observed that the CTMRG scheme converges better in the case of highly-entangled systems, as for example in the metal phase of the model \eqref{eq:ttV} close to the phase transition. In gapped systems both methods seem to converge equally well, with the directional CTM yielding slightly better accuracies than the present scheme. We used the former to address the gapped phase of the model \eqref{eq:free}, and the latter in all other simulations.

\end{document}